\documentclass{revtex4}

\usepackage{array}
\usepackage{graphicx}
\usepackage{hyperref}

\DeclareGraphicsExtensions{.eps}

\begin{document}

\title{All-optical reshaping of light pulses using $\chi^{(2)}$ media}

\author{Kuanshou Zhang, Laurent Longchambon, Thomas Coudreau
  \email{coudreau@spectro.jussieu.fr}, Claude Fabre}

\affiliation{Laboratoire Kastler Brossel, Universit\'e Pierre et Marie
  Curie, Campus Jussieu, Case 74, 75252 Paris cedex 05, France}

\date{\today}

 \begin{abstract} We have developed a new method based on 
two  cavities containing $\chi^{(2)}$ media to reshape optical pulses by an 
all-optical technique. The system is entirely passive \emph{i.e.}, all the energy is 
brought by the incoming pulse and uses two successive optical cavities with 
independent thresholds. The output pulse is close to a rectangular
shape. We show that this technique could be extended to high bit rates
and telecommunication wavelength using very small cavities containing
current nonlinear materials.
\end{abstract} 

\maketitle

\section{Introduction}

In optical telecommunications, multiple amplification and attenuation
of information-carrying light pulses leads to an amplification of
noise which quickly deteriorates the pulse shape, and therefore
increases the Bit Error Rate. To overcome such a signal distortion, an
active technique of pulse regeneration is used, which is usually
performed by opto-electronic techniques. The so called 3R regeneration
implies a Re-amplification, Re-shaping and Re-timing of the pulses. An
all-optical method to perform these different regeneration functions
would be potentially faster and with broader bandwidths. Several
methods have been proposed so far and are actively studied, based on
third-order nonlinearities or nonlinear amplifiers
\cite{regeneration,regeneration2} to implement this function.

Second order nonlinearities have not been studied so far to implement
regeneration functions, in spite of the high nonlinear effects that
they are likely to produce even at low input powers, and of their
intrinsic ultrashort response time. We propose here a passive and
efficient method for reshaping optical pulses which uses two
successive optical cavities containing nonlinear $\chi^{(2)}$
crystals. We give the results of a first experiment, which show that
our proposed scheme is actually able to reshape low power light
pulses, performed at $1.06 \mu m$ and at low bit rates. We also
discuss the potentialities of extension of our technique to real
optical telecommunication conditions.

Our scheme consists of two non-linear optical cavities which have
input-output characteristics with a threshold behavior, and lead to a
reshaping of the input pulse. The first cavity has a minimum
threshold~: as in a laser or an OPO, no significant power exits the
cavity below a certain value, $P^{low}_{threshold}$ of the pump power.
As a result, below $P^{low}_{threshold}$ the transmitted power is
zero. The second cavity has a maximum threshold
$P^{high}_{threshold}$~: an input signal with a power above
$P^{high}_{threshold}$ is transmitted with a constant value equal to
$P^{high}_{threshold}$. The total transfer function of the two optical
devices put in series is thus a steep Heavyside step function~: with
an incident distorted pulse, the output pulse is close to a square
pulse.

We begin in section \ref{sec:calculs} by giving the theoretical
expressions of the transfer functions of the two cavities. Section
\ref{sec:setup} is devoted to the description of the experimental
set-up while in section \ref{sec:results} we describe the experimental
results obtained at the wavelength of $1.06~\mu m$. Finally, in
section \ref{sec:perspective} we analyze the possibility to apply
these ideas to pulses having a wavelength of $1.5~\mu m$ and to very
high bit rates.

\section{Theoretical analysis \label{sec:calculs}}

\subsection{First optical system : the intracavity SHG/OPO}

The first cavity contains a type II $\chi^{(2)}$ crystal, and is
illuminated by a beam at frequency $\omega_0$, polarized linearly at
$+45^\circ$ of the crystal axes. The cavity is assumed to be at the
same time resonant for the ordinary and extraordinary waves at the
input frequency $\omega_0$, as well as for its second harmonic
$2\omega_0$, and to have a single coupling mirror at both frequencies.
The well-known following equations \cite{debuisschert} describe the
nonlinear resonant coupling between the intracavity amplitudes of the
three interacting fields in the steady-state regime and in the case of
exact triple resonance :
\begin{eqnarray}
 (1-r) A_1 &=& g A_0 A_2^\ast  + t \frac{1}{\sqrt{2}} A^{in}\label{eq:shgopo1}\\
 (1-r) A_2 &=& g A_0  A_1^\ast  + t \frac{1}{\sqrt{2}} A^{in}\label{eq:shgopo2}\\
 (1-r_0) A_0 &=& -g A_1 A_2\label{eq:shgopo3}
\end{eqnarray} where $A_1$ and $A_2$ are the intracavity ordinary and extraordinary 
wave envelope amplitudes at the fundamental frequency $\omega_0$,
$A_0$ the intracavity second harmonic amplitude and $A^{in}$ the input
field amplitude, $r$ and $t$ are the amplitude reflection and
transmission coefficients of the cavity coupling mirror at the
fundamental frequency, $r_0$ and $t_0$ the same quantities at the
second harmonic frequency. These equations are simplified if one
introduces the rotated basis: \[ A_\pm = \frac{1}{\sqrt 2} (A_1 \pm
A_2) \] which represent the amplitude of the fundamental field along
the two directions at $\pm 45^\circ$ from the crystal extraordinary
and ordinary directions. The steady state equations can now be written
in a decoupled form
\begin{eqnarray}
(1-r)A_+ &=& - g A_0 A_+^* + t A_+^{in} \\
(1-r)A_- &=&  + g A_0 A_-^* \label{eq:Amoins} \\
(1-r_0)A_0 &=& -\frac{g}{2} (A_+^2 - A_-^2)
\end{eqnarray}

These equations are readily solved \cite{ou}. One finds that there are
two different regimes, depending on the input intensity
$P^{in}=|A^{in}|^2$, separated by a "pitchfork bifurcation" occurring
at an intensity $P^{low}_{threshold}$ \cite{pitchfork} ( see figure
(\ref{fig:theotransferfn1}))~:
\begin{itemize}
\item For $P^{in} < P^{low}_{threshold}$, one has 
\begin{eqnarray} 
A_- &=& 0 \\ 
\left(1-r-\frac{g^2}{2(1-r_0)}|A_+|^2\right) A_+ &=& t A_+^{in}
\label{eq:belowthresh} \\ 
A_0 &=& -\frac{g}{2(1-r_0)} A_+^2 
\end{eqnarray}
\begin{figure} 
\centerline{\includegraphics[width=15cm,clip=]{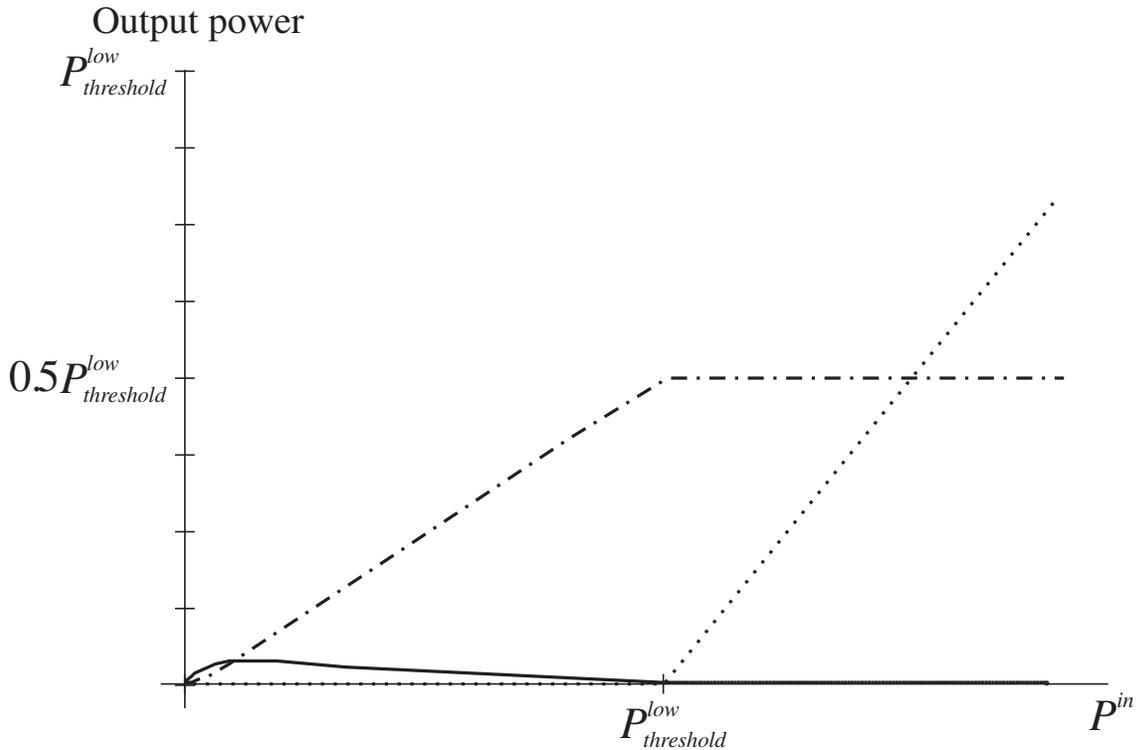}} 
\caption{Intracavity pump power along the + polarization $|A_+|$ (continuous 
  line), intracavity pump power along the - polarization, $|A_-|^2$
  (dotted line) and intracavity second harmonic power, $|A_0|^2$ (dash
  dotted line) for the first cavity as a function of the input pump
  power.\label{fig:theotransferfn1}} 
\end{figure}
This is the usual behavior of an intracavity doubly resonant frequency
doubler, in which more and more pump power is converted into second
harmonic as the pump power increases. \item For $P^{in} >
P^{low}_{threshold}$, one has $A_-\neq 0$, and :
\begin{eqnarray}
A_+ &=& \frac{t}{2(1-r)}A_+^{in} \\
A_-^2 &=& \frac{1+r}{4(1-r)}(A_+^{in})^2-\frac{2(1-r)(1-r_0)}{g^2} \label{eq:abovethresh} \\
A_0 &=& \frac{1-r}{g} 
\end{eqnarray} 
The intracavity second harmonic power is sufficient to generate an
oscillation on the $A_-$ mode. The behavior of the system is similar
to an Optical Parametric Oscillator : the second harmonic field is now
clamped to its threshold value while the $A_-$ power increases
linearly with the input power. If one uses as the output of the device
the output field orthogonally polarized to the input field, and at the
same frequency, $A_-^{out}= t A_-$, one will have a zero output below
threshold, and a value rapidly growing with the input above threshold,
which is the behavior that we need in our reshaping device. It is
worthy to note that the present threshold $P^{low}_{threshold}$, equal
to $ \frac{T^2 T_0}{2 g^2}$ for small intensity transmission
coefficients at both frequencies, $T=t^2$ and $T_0=t_0^2$, is the
threshold of a triply resonant OPO, which can be in the mW range in
optimized conditions \cite{2microns}.
\end{itemize}

Two difficulties must be solved to operate this device. First, we have
assumed perfect triple resonance to solve the equations. This is not
an easy condition to fulfill, as the fields $A_0$, $A_1$ and $A_2$ see
three different indices in the type II nonlinear crystal,
corresponding to three different optical paths. As all the frequencies
are fixed, we must then adjust precisely three parameters of the
system to fulfill the triple resonance conditions. Two parameters are
easily controllable : the cavity length and the crystal temperature,
but a third one is needed. This last one could be the setting of a
variable birefringent system in the cavity. We chose another approach,
which is to insert a quarter-wave-plate (QWP) in the linear cavity,
adjusted so that it induces a $90^\circ$ rotation of the polarization
plane for the fundamental wave when it is crossed twice by the beam
inside the cavity, and does not change the second harmonic field
polarization. In this configuration, for any crystal temperature, the
eigenmodes of the cavity are automatically $(A_+,A_-)$. We need then
to adjust only two parameters to ensure the triple resonance condition
: the cavity length and the crystal temperature. Secondly, we have
assumed a degenerate operation for the OPO. This is one possibility
for the system. But a non-degenerate operation of the OPO can also
take place, with signal and idler modes oscillating at different
frequencies $\omega_1$ and $\omega_2$ such that~:
$\omega_1+\omega_2=\omega_0$. This regime of cascading has been
theoretically studied \cite{theoriecascading1,theoriecascading2} and
observed \cite{manipcascading}. Actually, the system will oscillate in
the regime (degenerate or non-degenerate) which has the lowest
threshold. We will see in section \ref{sec:manipopoautopompe} that one
can find conditions for which it is indeed possible to observe the
stable degenerate operation that is needed for reshaping without
frequency change, which is necessary for practical applications. A
forthcoming publication will give more precise insight into the
operation of this kind of optical devices that contain at the same
time second order nonlinear elements and birefringent elements.

\subsection{Second optical system : a non-degenerate OPO}

The second system consists of a standard, non-degenerate, triply
resonant OPO. It is well known that, in a laser above threshold,
the gain is clamped to its threshold value by the condition that
the saturated gain must equal the losses in the steady-state
regime. The same behavior occurs in an OPO, as can be seen very
easily from its steady state equations, which are very similar to
eqs (\ref{eq:shgopo1}-\ref{eq:shgopo3}), except that now the input power is put in the
cavity on the high frequency mode :
\begin{eqnarray}
 (1-r_s) A_s &=& g A_p A_i^\ast\\
 (1-r_i) A_i &=& g A_p  A_s^\ast\\
 (1-(r_p)^2) A_p &=& - g A_s A_i+ t_p A_p^{in}
\end{eqnarray} 
where $A_s$ and $A_i$ are the intracavity signal and idler mode
amplitudes (these two modes oscillating at different frequencies),
$A_p$ the intracavity pump amplitude, and $A_p^{in}$ the input pump
amplitude; $r_s$, $r_i$ and $r_p$ are the amplitude reflection
coefficients of the coupling mirror at the signal, idler and pump
frequencies. For the pump field, one has assume that the cavity has
two identical coupling mirrors, of amplitude reflection and
transmission coefficients $r_p$ and $t_p$, one used as the input, the
second as the output of our optical device.

\begin{figure}[h]
 \centerline{\includegraphics[width=8cm,clip=]{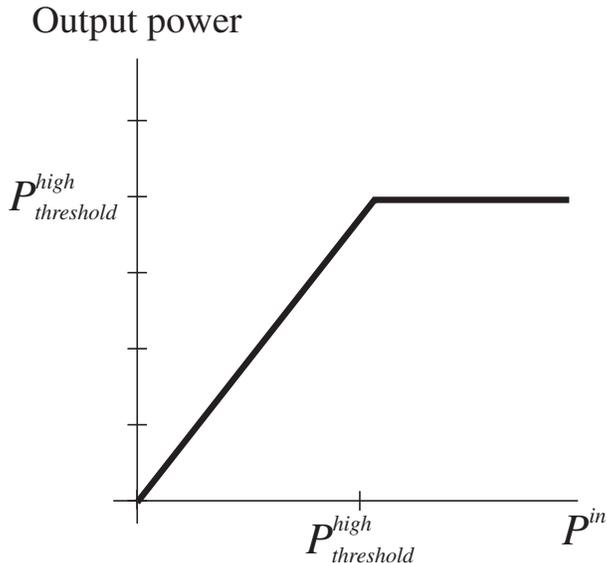}} 
\caption{Transmitted power $I_{transm.}$ by the second cavity as a
  function of the input pump power $I_{in}$.
  \label{fig:theotransferfn2}} \end{figure}

Solving these very well-known equations \cite{debuisschert}, one finds that there are
two different regimes, depending on the pump intensity
$P_{pump}=|A_p^{in}|^2$, separated by a threshold
$P^{high}_{threshold}$ (see figure (\ref{fig:theotransferfn2})):
\begin{itemize} 
\item For $P_{pump} < P^{high}_{threshold} = \frac{t_p^2
    (1-r_s)(1-r_i)}{g^2}$, one has $A_s=A_i=0$. No parametric
  oscillation takes place.  The OPO cavity is a pure passive,
  resonant, Fabry-Perot cavity with input and output mirrors of equal
  transmission. Its transmission is therefore 1, and the device is
  exactly transparent.
\item For $P_{pump} > P^{high}_{threshold}$, one has $A_s\neq 0$ and
  $A_s\neq 0$, which occurs only when
  $|A_p|^2=\frac{(1-r_s)(1-r_i)}{g^2}$, whatever the input pump field
  is : the intracavity pump power is therefore clamped to a value
  independent of the input, and the output pump field is then clamped
  to its value at threshold. The excess power brought by the pump is
  then transferred to the signal and idler beams.
\end{itemize}

Let us stress that here also, the pump threshold
$P^{high}_{threshold}$ is the threshold of a triply resonant OPO,
which can be very low. But now the system is much simpler to operate
than the previous one, as the frequencies of the signal and idler
modes are not a priori given (except that their sum is equal to the
pump frequency). There is one more degree of freedom than in the first
device, and the cavity length and crystal temperature are the only two
parameters that need to be adjusted to get the triply resonant
condition.

\subsection{Total system}

\begin{figure} \centerline{\includegraphics[width=8cm,clip=]{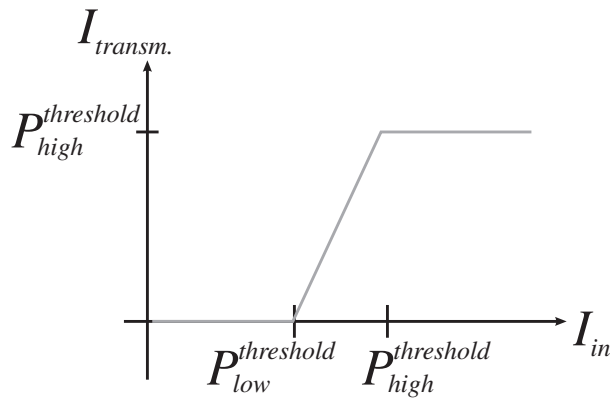}} 
\caption{Transmitted power through the total systems as a function of the 
 input pump power. \label{fig:theotransferfntot}} \end{figure}

If the two previously described devices are put in series, one obtains
an overall input-output characteristics which is sketched in figure
(\ref{fig:theotransferfntot}).  This curve is close to the one we need
for pulse reshaping, except for the intermediate region
$P^{low}_{threshold} < P_{pump} < P^{high}_{threshold}$, for which the
response is linear. The best reshaping will be obtained when this
central part is as steep as possible.

\section{Experimental set-up \label{sec:setup}}

\begin{figure} \centerline{\includegraphics[width=15cm,clip=]{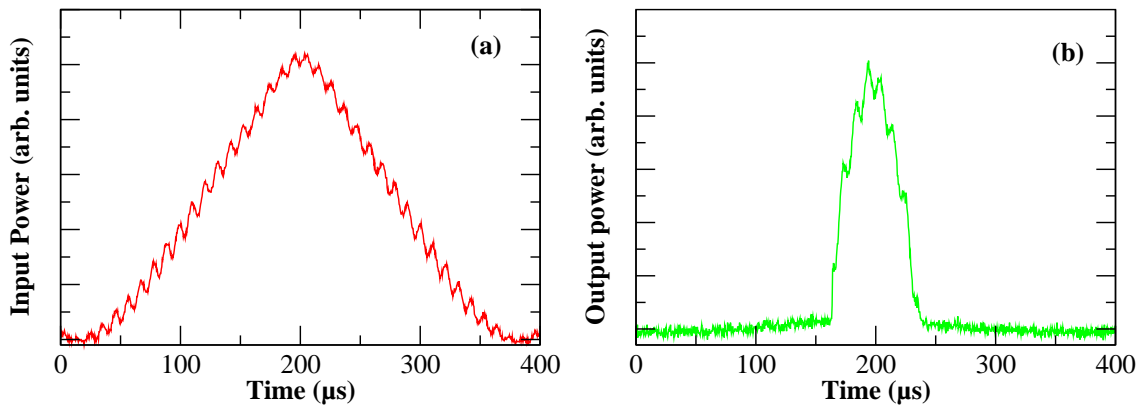}} 
\caption{Experimental results for cavity 1
  \label{fig:intensitycavity1}~: (a) input power as a function of time
  (b) output power as a function of time} \end{figure}

The input beam is produced by a Nd:YAG laser at $1.06~\mu m$
(Lightwave 126-1064-700). In order to produce a light pulse, we use an
acousto-optic modulator to modulate the intensity of the transmitted
beam (modulation frequency $\approx 3~kHz$). To mimick the high
frequency noise existing on the pulse, we superimpose on the enveloppe
a high frequency modulation ($\approx$ 100~$kHz$) and simulate the
high frequency noise (fig.~\ref{fig:intensitycavity1}.a).

\subsection{Intracavity SHG/OPO}

The characteristics of the two cavities are summarized in
Tab.\ref{tab:characteristics}. The first cavity has highly reflecting
mirrors both for $1.06~\mu m$ and $0.53~\mu m$. The pump at $1.06~\mu
m $ is sent at a $45^\circ$ angle with respect to the crystal axis.
The output beam is separated from the input by using a polarizing beam
splitter in front of the optical device. At low input power, it is
first converted in green light by a standard doubling process. When
the intracavity green power is sufficient, parametric down conversion
occurs which transfers the power back to the pump wavelength but on
the orthogonal polarization. However, this system has a high
threshold, above $600~mW$ in our experimental conditions. In order to
reduce his threshold, we have added a quarter waveplate at $1.06~\mu
m$ inside the cavity. In that case, we observed a much more efficient
intracavity second harmonic generation, and a parametric oscillation
threshold could be as low as 300~$mW$. This fairly large threshold is
mainly due to the fact that the reflection coefficients were not
optimized (in particular, the green reflection coefficients should be
maximum which is not the case).

\begin{table} \caption{Characteristics of the cavities~: $R_c$ radius of curvature in 
mm, R$_{(2)\lambda}$ reflection coefficient at $(2)\lambda$.\label{tab:characteristics} }
\begin{tabular}{|m{2.2cm}|m{1.5cm}|m{2.5cm}|m{.7cm}|m{1.4cm}|m{1.4cm}|m{.7cm}|m{1.4cm}|m{1.4cm}|}
\hline & \makebox[1.5cm]{\raisebox{-2ex}[0cm][0cm]{Crystal}} &
\makebox[2.5cm]{\raisebox{-2ex}[0cm][0cm]{ \begin{tabular}{c}Cavity length
\\($mm$)\end{tabular}}} &
\multicolumn{3}{c|}{\makebox[3.5cm]{Input Mirror}}& \multicolumn{3}{c|}{\makebox[3.5cm]{End Mirror}} \\
\cline{4-9} &  & & \makebox[.7cm]{$R_c$} & \makebox[1.4cm]{R$_{\lambda}$}& 
\makebox[1.4cm]{R$_{2\lambda}$} & \makebox[.7cm]{$R_c$} & 
\makebox[1.4cm]{R$_{\lambda}$}& \makebox[1.4cm]{R$_{2\lambda}$} \\ \hline
\makebox[2.2cm]{\begin{tabular}{c}%
Cavity 1\\%
$\lambda=532~$nm%
\end{tabular}}
 &\makebox[1.5cm]{KTP} &\makebox[2.5cm]{50} & \makebox[.7cm]{50} & \makebox[1.4cm]{$>$99.9\%} & \makebox[1.4cm]{95\%} & \makebox[.7cm]{50} & \makebox[1.4cm]{90\%} & \makebox[1.4cm]{$>$99.9\%}
 \\
\hline
\makebox[2.2cm]{\begin{tabular}{c}%
Cavity 2\\%
$\lambda=1064$~nm%
\end{tabular}}
 & \makebox[1.5cm]{PPLN} & \makebox[2.5cm]{65} & \makebox[.7cm]{30} &
 \makebox[1.4cm]{87\%} & \makebox[1.4cm]{99.8\%} &\makebox[.7cm]{30} &
 \makebox[1.4cm]{99.8\%} & \makebox[1.4cm]{99\%} \\
\hline \end{tabular} \end{table}

\subsection{Non-degenerate OPO}

This system has been described extensively in a previous publication
\cite{2microns,sqpompe}. Let us briefly mention here its main features
: A PPLN crystal with an inversion period of 30~$\mu m$ is placed
inside a symmetric cavity which has a large finesse (over 200) for the
wavelengths around $2~\mu m$ and a lower finesse at $1~\mu m$ (around
40). The crystal temperature is kept close to the degeneracy
temperature (parametric conversion between $1.06~\mu m$ and $2.12~\mu
m$) so that parametric oscillation occurred for all cavity lengths
within a pump resonance, because of the overlap between the
oscillation ranges of nearby signal and idler pairs of modes of
wavelengths close to $2~\mu m$ \cite{2microns}. The threshold was on
in the order of a few $mW$. This threshold was chosen to obtain a
transmitted field as close as possible to a square function~: its
value is much lower than the maximum intensity at the output of the
first cavity to ensure a steep transmission function. Let us also
mention that in our experiment, the input and output mirrors for the
$1.06~\mu m$ beam where not of equal transmission. As a result the
power of the transmitted beam was very small compared to the input
one, and not equal, as in the theoretical approach of the previous
section. This is due to the fact that we used, instead of an optimized
cavity, an existing one as used in the experiment described in ref
\cite{2microns}.

\section{Experimental results\label{sec:results}}

\subsection{Intracavity SHG/OPO \label{sec:manipopoautopompe}}

\begin{figure}
 \centerline{\includegraphics[height=6cm,clip=]{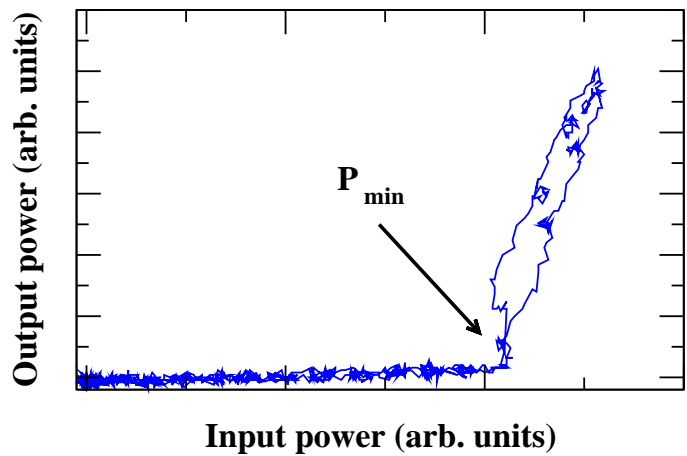}}
\caption{Transfer function of the first
  cavity\label{fig:exptransferfn1}} 
\end{figure}

Fig. \ref{fig:intensitycavity1} shows the input and output intensities
of the first device as a function of time. The maximum input power on
this cavity was 350~$mW$ in order to be above the threshold
$P^{low}_{threshold}$ mentioned above. It can be seen on fig.
\ref{fig:exptransferfn1} that the effect of the cavity is close to a
perfect "high-pass filter", as far as intensities are concerned~:
powers below $P^{low}_{threshold}$ are not transmitted while those
above this value are linearly transmitted.

\begin{figure}
\centerline{\includegraphics[width=0.4\textwidth,clip=]{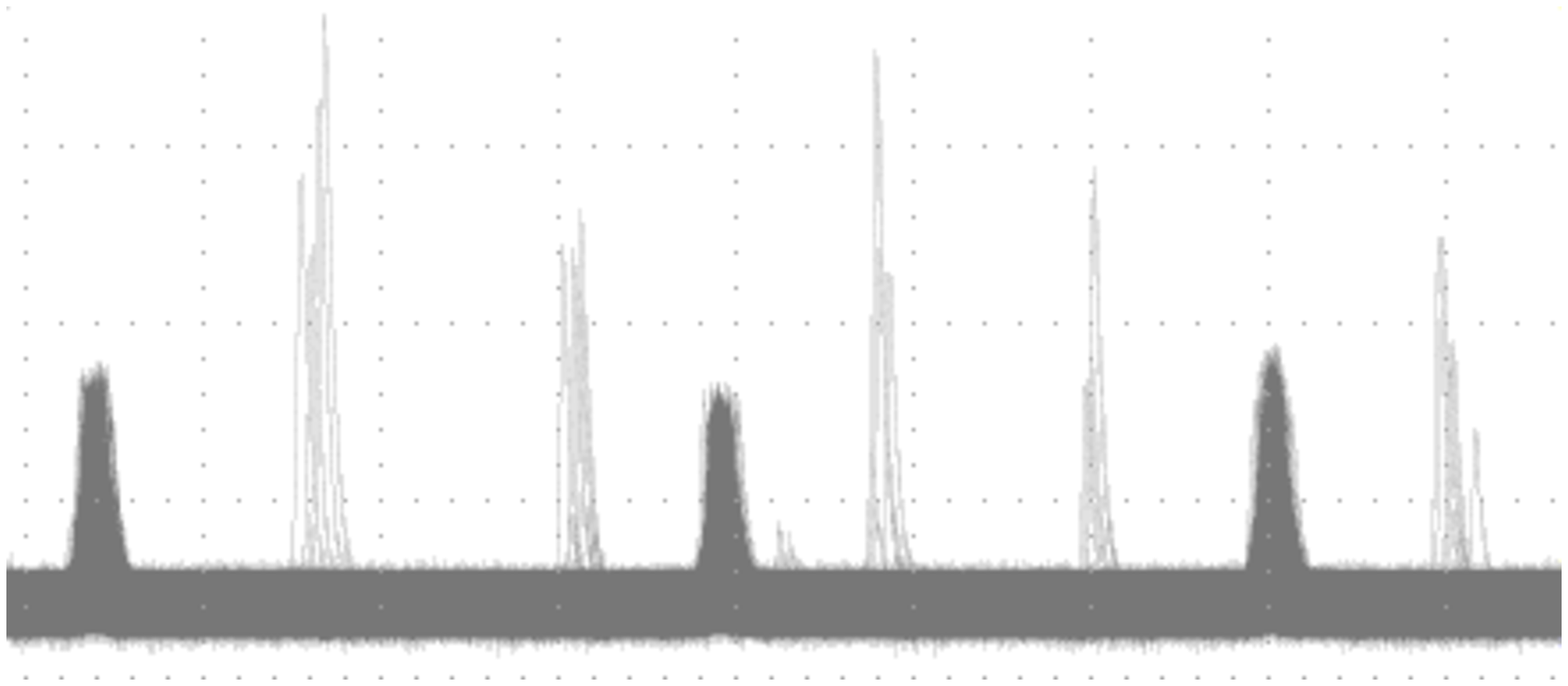}%
  \hspace{.5cm}\includegraphics[width=0.4\textwidth,clip=]{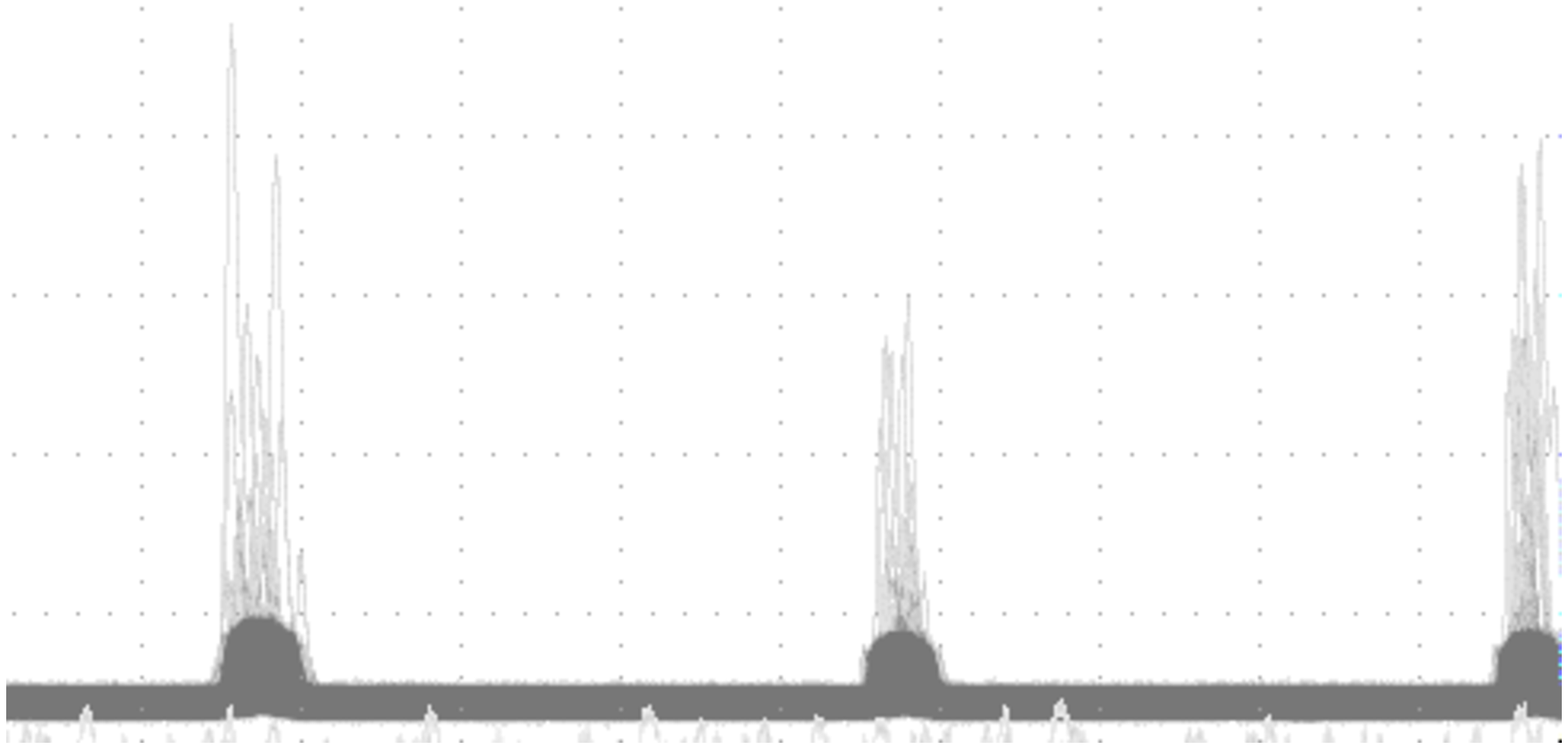}}
\caption{Down converted beam and pump beam transmitted intensities
  through the confocal Fabry-Perot cavity as a function of the
  confocal Fabry-Perot cavity length. The sharp and light peak is the
  transmission of the down converted output and the blunt and dark
  peak is the transmission of the pump beam. Left is the
  non-degenerate case while right shows the degenerate
  case\label{fig:checkdegen}} \end{figure}

A crucial point of the device is that the output beam (orthogonally
polarized reflected pump beam) is at the same frequency as the input
beam. In a first experiment, we looked at the interference pattern
between the input and the output beam, and observed that there were no
fringes when the second harmonic beam was not $TEM_{00}$. Clear
fringes only appeared when, by a careful alignment and crystal
temperature tuning, the green output was $TEM_{00}$. The frequency
degenerate operation was corroborated by a second experiment, where we
used a confocal Fabry-Perot cavity to monitor the frequency of the
output beam, orthogonally polarized with respect to the input beam.
The confocal Fabry-Perot cavity is formed by two curved mirrors with
radius of 50~$mm$ and reflectivity of 95\% for $1.064~\mu m$ that the
free spectral range is 1.5~$GHz$. Figure \ref{fig:checkdegen} shows
the output beam and the pump beam transmitted intensities through the
confocal Fabry-Perot cavity when scanning the analysis cavity at
$60~Hz$ and scanning the self pumped OPO at 650Hz~: this figure shows
that, when the system is not perfectly tuned up, it oscillates in a
non-degenerate regime and generates sidebands around the pump
frequency (fig.  \ref{fig:checkdegen}, left), whereas one can find
experimental conditions for which the down-converted output has the
same frequency as the pump, within the experimental uncertainties
(fig. \ref{fig:checkdegen}, right).

\subsection{Non degenerate OPO}

\begin{figure} \centerline{\includegraphics[height=6cm,clip=]{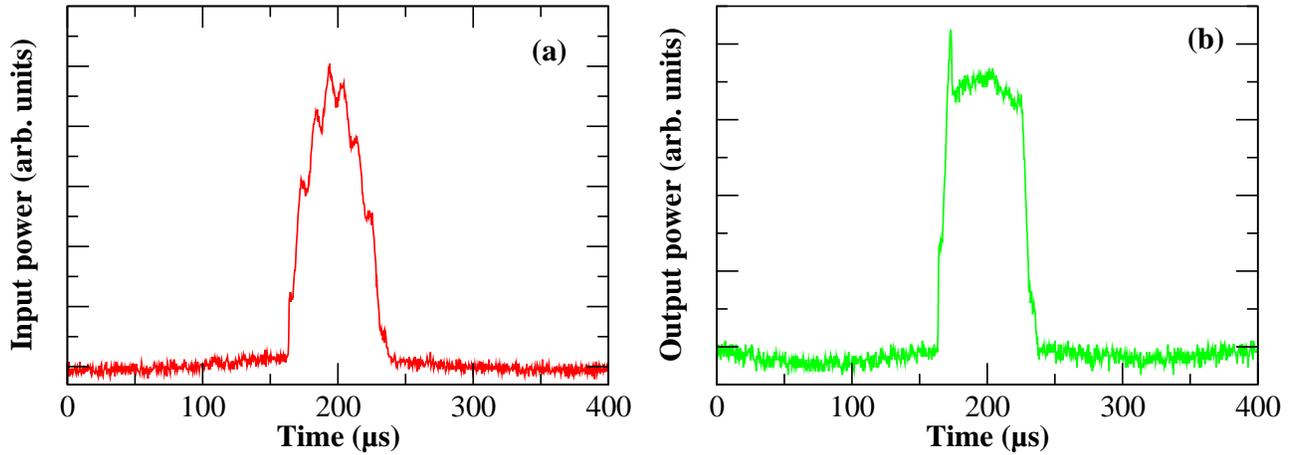}}
\caption{Experimental results for cavity 1
  \label{fig:intensitycavity2}~: (a) input power as a  
  function of time (b) output power as a function of time}
\end{figure}

\begin{figure} \centerline{\includegraphics[height=6cm,clip=]{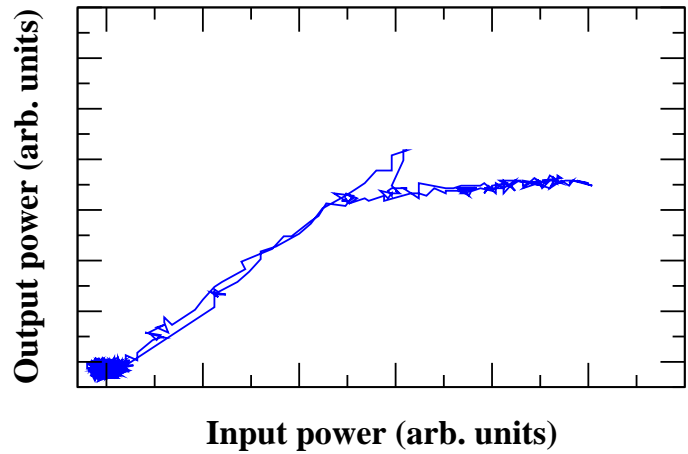}}
\caption{Transfer function of the second cavity
  \label{fig:exptransferfn2}} \end{figure}

We have plotted on fig. \ref{fig:intensitycavity2} the relevant
intensities for the second cavity as a function of time~:
\ref{fig:intensitycavity2}.a shows the incident intensity, while
\ref{fig:intensitycavity2}.b shows the transmitted intensity. The
output intensity displays a very clear clamping of the power above the
threshold $P^{high}_{threshold}$, at a value equal to the transmitted
pump power at threshold, typically a few $mW$. On fig.
\ref{fig:exptransferfn2}, the "low-pass filter" (for intensities)
effect of this cavity is shown via its transfer function.


The peak that one observes at the beginning of the flat top of the
transmitted intensity is due to a dynamical effect of delayed
bifurcation that has already been observed in OPOs when their pump
power is scanned with time \cite{richy}. As the incoming pump power
increases above threshold, the onset of the oscillation is delayed by
a time interval that is larger than the characteristic evolution times
of the cavity.

\subsection{Total cascaded system}

\begin{figure} \centerline{\includegraphics[height=5cm,clip=]{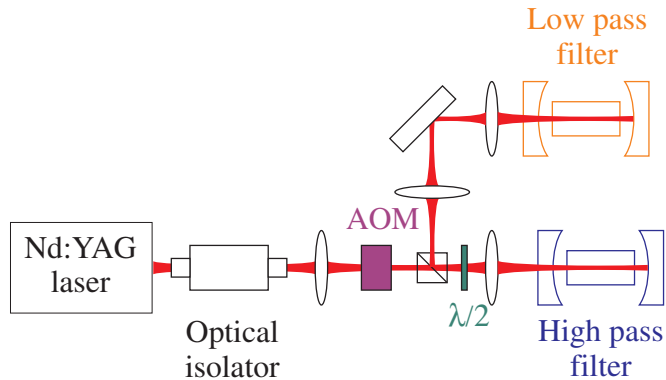}}
  \caption{Experimental set-up with the two cavities in series showing the intracavity SHG/OPO (low pass filter) and non degenerate OPO (high pass filter) \label{fig:totalsetup}}
\end{figure}

\begin{figure} \centerline{\includegraphics[height=5cm,clip=]{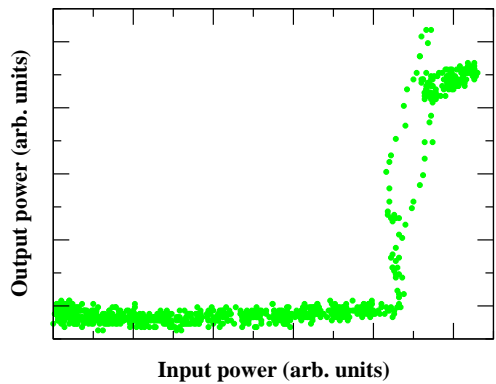}}
  \caption{Total transfer function of the system \label{fig:exptransferfntot}}
\end{figure}

The complete experimental set-up is shown on fig \ref{fig:totalsetup}.
The beam reflected by the intracavity SHG/OPO is separated from the
input beam by a polarizing beam splitter and sent to the
non-degenerate OPO.  The output beam intensity and the experimental
transfer function for the total system are shown on fig.
\ref{fig:exptransferfntot}. One observes that the time dependence of
the output beam intensity is close to a rectangular pulse, and
accordingly that the transfer function shows the desired behavior with
a response very close to a step function.


\section{Possible implementation to very high bit rates at $1.5 \mu m$
  \label{sec:perspective}} 

In all-optical pulse reshaping systems, two parameters play an
important role : the operating power and the response time. In the
present demonstration experiment, the incident power on the first
cavity is about 350~$mW$, which is beyond the current powers of
optical telecommunication pulses. This is mainly due to the fact that
the first cavity is not optimized. Thresholds in the range of tens of
$mW$ could be obtained by using high quality materials and
coatings and optimizing the cavity parameters. The response time of
our present system is rather long, of the order of 200 $ns$~: it is
related to the cavity build-up time of the two cavities which are long
cavities with rather high finesses. The most important question for
the future of our proposed technique is whether our system can be
modified and optimized in order to be able to reach very high bit
rates, namely 40~$Gbit/s$, and at the telecommunication wavelength of
$1.5~\mu m$. We will address this question in the present section.

Several nonlinear materials are compatible with the telecommunication
wavelength, namely Gallium Arsenide, Aluminium Gallium Arsenide, or
Zinc Selenide. These materials posses very large nonlinearities, as
large as $120~pm/V$ in the case of Gallium Arsenide but they cannot be
phase matched using birefringence.

The rising time issue is the most difficult to solve. In order to
reduce this time, one must use small cavities, and therefore shorter
non-linear crystals, and/or lower reflectivities for the mirrors, two
methods which have the detrimental effect of increasing the thresholds
at the same time.

The threshold of a triply resonant OPO is given by 
\begin{equation}
P_{threshold} = \frac{T_0 T_1 T_2}{64 \chi ^2 L_c ^2} 
\end{equation}
where $T_i$ denotes the transmission coefficient of the cavity for
mode $i$, $L_c$ the crystal length and $\chi$ is the non linear
coupling strength depending on the nonlinearity, geometry and optical
indices through the relation~:
\begin{equation}
  \chi = d_{eff} \frac{w_0 w_1 w_2}{w_0^2 w_1 ^2 + w_0^2 w_2^2 + w_1^2
    w_2^2} \sqrt{\frac{\hbar \omega_0
      \omega_1 \omega_2}{\pi \epsilon_0 c^3 n_0 n_1 n_2}}
\end{equation}
where $d_{eff}$ is the crystal nonlinearity in $m/V$, $w_i$ is the
waist size, $\omega_i$ is the pulsation and $n_i$ the indices of
refraction of the three interacting modes. Assuming the system is
operated close to frequency degeneracy, that is $\omega_{1,2} =
\omega_0/2$ and that the pump, signal and idler indices and Rayleigh
ranges are identical, the nonlinear coupling strength can be expressed
in terms of wavelength and Rayleigh length, as~:
\begin{equation}
  \left( \frac{\chi_{exp}}{\chi} \right)^2=  \left(
    \frac{(d_{eff})_{exp}}{d_{eff}}\right)^2 
  \frac{z_R}{(z_R)_{exp}} \left( \frac{\lambda_0 }{(\lambda_0)_{exp}}
  \right)^4 \left( \frac{n}{n_{exp}} \right)^2
\end{equation}
where the index $exp$ denotes experimental values, $\lambda_0$,
$(\lambda_0)_{exp}$ denote the pump wavelength in the vacuum, these
two wavelengths being different due to the fact that our experiment was
performed with a $(\lambda_0)_{exp}=532~nm$ pump while a telecom
implementation would require $\lambda_0 \approx 1550~nm$.

If we also assume that the cavity length is equal to the crystal
length, one can use the expression for the finesse as a function of
the cavity rise time
\begin{equation}
  \mathcal F = \frac{\pi c \tau}{2 n L_c}
\end{equation}
We now obtain the following formula for the  ratio of the thresholds~:

\begin{equation}
\frac{P_{threshold}}{P_{exp}} = (\mathcal F_0 \mathcal F_1
    \mathcal F_2)_{exp}\left(\frac{2 n}{
    \pi c \tau } \right)^3
\left(\frac{(d_{eff})_{exp}}{d_{eff}}\right)^2
\frac{z_R}{(z_R)_{exp}}  \left( \frac{
    \lambda_0}{(\lambda_0)_{exp}} \right)^4 \left( \frac{n}{n_{exp}}
\right)^2  L_{exp}^2 L_c 
\end{equation}

One obtains a counterintuitive result, namely that the threshold
becomes lower as the crystal/cavity length diminishes. This is due to
the fact that the finesse is fixed by the value of the cavity length
in order to keep the cavity lifetime $\tau$ constant. A small cavity
is thus desirable to obtain a low threshold as well as a short rise
time. The experimental values correspond to an optimized cavity
containing a KTP crystal pumped by a frequency doubled Nd:YAG laser at
532$~nm$ generating signal and idler around 1064$~nm$ yielding
$P_{exp}=20~mW$ with $n_{exp} \approx 1.8$, $(d_{eff})_{exp}=3 ~pm/V$,
$L_{exp}=1~cm$, $ (\mathcal F_{0,1,2})_{exp} = 115$. When one sets a
cavity/crystal length of 10$~\mu m$, a rise time $\tau$ of $10~ps$
(compatible with operating bit rate of 40 GBit/s), and a Rayleigh
length of 10$~\mu m$ compatible with the crystal length, one obtains a
finesse $\mathcal F \approx 140 $ and threshold $P_{threshold} \approx
110~mW$ with the values corresponding to gallium arsenide ($n=3.4$ and
$d_{eff} = 120~pm/V$). This value of the threshold is compatible with
the values used in optical telecommunications. The very short length
of the crystal is also an advantage as one can have a crystal length
equal to or shorter than the coherence length of the material which
partly removes the problem of phase-matching. It is important to note
that the variation of the threshold with $\tau$ is very fast so that a
small increase of the rise time up to 15$~ps$ leads to a similar
threshold using Lithium Niobate ($n=2.2$ and $d_{eff} = 20~pm/V$).

We have shown in this section that the realization of very short rise
time systems for all optical reshaping is within reach by developing
current techniques and using available materials. However, because of
the presence of the resonance cavity, this system is only able to
reshape optical pulses centered around a given wavelength.

\section{Conclusion}

We have demonstrated that optical cavities containing $\chi^{(2)}$
media can be used for all optical passive reshaping of optical pulses.
We have experimentally obtained reshaping with a threshold compatible
with optical powers propagating in optical fibers. We have shown that
very fast response times can be realized using very short monolithic
cavities made of high nonlinearity crystals.

\begin{acknowledgments}
  Laboratoire Kastler-Brossel, of the Ecole Normale Sup\'{e}rieure and
  the Universit\'{e} Pierre et Marie Curie, is associated with the
  Centre National de la Recherche Scientifique.  This work was
  supported by France-Telecom (project CTI n$^\circ$ 98-9.003).  Zhang
  Kuanshou was on leave from the Key Laboratory for Quantum Optics,
  Shanxi University, Taiyuan, China.  T. Coudreau is also at the
  P\^ole Mat{\'e}riaux et Ph{\'e}nom{\`e}nes Quantiques FR CNRS 2437.
  The authors want to thank V. Berger for fruitful discussions.
\end{acknowledgments}

\end{document}